\documentclass[conference,compsoc]{IEEEtran}
\IEEEoverridecommandlockouts
\usepackage[T1]{fontenc}
\usepackage[utf8]{inputenc}

\usepackage{amsmath,amssymb,amsthm}
\usepackage{newtxmath}
\usepackage{inconsolata}

\usepackage{algorithm}
\usepackage[noend]{algpseudocode}

\usepackage{xspace}
\usepackage{multirow}
\usepackage{multicol}

\usepackage{makecell}
\usepackage{thmtools, thm-restate}

\usepackage{enumitem} 
\setlist{leftmargin=*} 

\usepackage[dvipsnames]{xcolor}
\usepackage{soul}
\usepackage{minted}
\usepackage{comment}
\usepackage{booktabs}
\setminted{
    linenos,          
    fontsize=\small,  
    breaklines,       
    xleftmargin=15pt,   
}

\newif\ifdraft 
\drafttrue

\newif\ifrevisiondiff
\revisiondifffalse

\newcommand{\soohyuktext}[1]{\ifdraft{\color{olive}#1}\else#1\fi}
\newcommand{\ternop}{\;\mathbin{?}\;}
\newcommand{\terncol}{\;\mathbin{:}\;}

\ifdraft\fi

\usepackage{graphicx} 
\usepackage{subcaption} 
\usepackage{adjustbox}

\usepackage{listings}

\usepackage[dvipsnames]{xcolor}

\lstdefinestyle{mystyle}{
  basicstyle=\ttfamily\footnotesize,
  keywordstyle={[1]\color{magenta}},
  keywordstyle={[2]\color{BrickRed}},
  keywordstyle={[3]\color{RoyalBlue}},
  commentstyle=\color{OliveGreen},
  stringstyle=\color{YellowOrange},
  numbers=left,
  numberstyle=\tiny\color{gray},
  numbersep=3pt,
}
\lstdefinelanguage{mylang} {
    morekeywords={[1]module, method},
    morekeywords={[2]Bool, HistoryT, Addr, Data},
    morekeywords={[3]if},
    sensitive=false,
    morecomment=[l]{//},
}
\usepackage[normalem]{ulem}

\renewcommand{\paragraph}[1]{%
  \vspace{5pt}\noindent\textbf{#1}\quad
}

\makeatletter
\newcommand{\newlineauthors}{%
  \end{@IEEEauthorhalign}\hfill\mbox{}\par
  \mbox{}\hfill\begin{@IEEEauthorhalign}
}
\makeatother

\usepackage{amsthm}
\usepackage{aliascnt}
\usepackage{bm}
\usepackage{subcaption}

\usepackage{hyperref}
\usepackage[nameinlink,capitalise,noabbrev]{cleveref} 

\usepackage{multirow, multicol}

\theoremstyle{plain}

\theoremstyle{definition}
\newtheorem{definition}{Definition}[section]
\newtheorem{example}{Example}[section]

\theoremstyle{remark}

\ifCLASSOPTIONcompsoc
  \usepackage[nocompress]{cite}
\else
  \usepackage{cite}
\fi
\ifCLASSINFOpdf
\else
\fi

\begin{document}

%


\title{Efficient Hardware Information-Flow Tracking for \\ Pre-Silicon Security Testing
}



\author{
\IEEEauthorblockN{Yu-Wei Fan\IEEEauthorrefmark{1}\thanks{\IEEEauthorrefmark{1} Both authors contributed equally to this research.}}
\IEEEauthorblockA{Princeton University\\
yf9172@princeton.edu}
\and
\IEEEauthorblockN{Yuheng Yang\IEEEauthorrefmark{1}}
\IEEEauthorblockA{MIT CSAIL\\
yuhengy@mit.edu}
\and
\IEEEauthorblockN{Christine Guo}
\IEEEauthorblockA{Princeton University\\
cg4302@princeton.edu}
\newlineauthors
\IEEEauthorblockN{\hspace{-1.5em}SooHyuk Cho}
\IEEEauthorblockA{\hspace{-1.5em}Princeton University\\
\hspace{-1.5em}soohyuk.cho@princeton.edu}
\and
\IEEEauthorblockN{\hspace{-2.5em}Thomas Bourgeat}
\IEEEauthorblockA{\hspace{-2.5em}EPFL\\
\hspace{-2.5em}Thomas.bourgeat@epfl.ch}
\and
\IEEEauthorblockN{\hspace{-2em}Mengjia Yan}
\IEEEauthorblockA{\hspace{-2em}MIT CSAIL\\
\hspace{-2em}mengjiay@mit.edu}
\and
\IEEEauthorblockN{\hspace{-2.5em}Sharad Malik}
\IEEEauthorblockA{\hspace{-2.5em}Princeton University\\
\hspace{-2.5em}sharad@princeton.edu}
}

\maketitle

\thispagestyle{plain}
\pagestyle{plain}

\begin{abstract}
Register-Transfer Level (RTL) simulation is widely used to test hardware before it is fabricated.
To allow testing for security related information flow properties, such as confidentiality and integrity, 
taint logic can be automatically added to the design to track how information flows through it.
However, taint logic instrumented by the state-of-the-art, such as CellIFT, 
makes simulation-based testing prohibitively expensive:
On our evaluation of Mega-BOOM (136K cells), it increases the instrumented design to 5.81$\times$ the original cell count and causes a 143.72$\times$ simulation slowdown.
The taint logic could be simplified to improve simulation speed, but it will inevitably trade off its precision.
This lightweight, imprecise taint logic will introduce false positives and may eventually result in even more overhead to check these false positives.


This paper explores the research question of where precision is actually needed in the design to overcome the overhead of false positives. It presents CEGAR-T, a framework that \emph{automatically} synthesizes taint logic that \emph{minimizes} the taint-logic instrumentation overhead while guaranteeing no false positives (relative to the precise CellIFT baseline). We have implemented CEGAR-T and evaluated it on the safe instruction set problem for timing side-channel security across open-source RISC-V cores. Over all evaluated cores, CEGAR-T reduces both instrumentation and simulation overhead, in geometric-mean, from 5.64$\times$ to 1.42$\times$ and from 34.65$\times$ to 1.79$\times$, respectively, without compromising the precision benefit of the CellIFT baseline.

\end{abstract}


%
\IEEEpeerreviewmaketitle

\section{Introduction}
\label{sec:introduction}








Modern hardware enforces security boundaries for privilege separation, trusted execution, confidentiality, and integrity, yet deployed attacks continue to show that subtle microarchitectural behavior can violate these boundaries~\cite{spectre,meltdown,foreshadow,zombieload,crosstalk}. 
Once such flaws reach silicon, they are difficult or too costly to eliminate completely with post-silicon patches, because mitigations are constrained by already-fabricated and deployed hardware~\cite{cellift,socSecuritySurvey,emulationSecuritySurvey}.
This makes pre-silicon hardware security testing a critical task, where at the RTL stage, internal design state is visible, security-relevant executions can be simulated, and vulnerable logic can still be fixed before fabrication~\cite{introspectre,specdoctor}.

Hardware information-flow tracking (IFT) via taint analysis~\cite{hift-survey}
is a practical workhorse for simulation-based pre-silicon security testing,
widely used in both academic work~\cite{glift,rtlift,cellift,hu2018property}
and industrial security-validation settings~\cite{opentitanSecurityVerification,Cycuity1,Cycuity2}.
It represents the target hardware security requirement as a source-to-sink
information flow property. 
The source represents secret data, untrusted input, or control state
whose influence should be restricted, while a sink represents protected design states or
behavior that should not be affected by that source under the target security property.
Given an information flow property, taint analysis instruments the RTL with taint bits and
taint logic, 
tracks how taint from the source propagates through the design, 
and flags a potential violation when taint reaches the sink.
We remark that taint analysis-based IFT is complementary to hardware fuzzing, which excels at exploring a large number of diverse executions~\cite{specdoctor,dejavuzz}, by providing a mechanism to determine whether those executions contain forbidden source-to-sink flows.
Taint logic is \emph{sound}: it over-approximates true information flow, 
whose semantics is given by the well-established non-interference (NI) semantics~\cite{non-interference}.
Its precision is determined by how tightly this approximation matches the NI semantics.
Prior work has shown that imprecise taint logic tends to over-taint and generate false positives, whereas more precise taint logic is often more complex, incurring prohibitively high simulation overhead~\cite{hift-survey}.
In the largest design in evaluation, 
Mega-BOOM~\cite{boom-v3} (136K cells), 
the state-of-the-art taint analysis tool CellIFT~\cite{cellift} increases the instrumented design to 5.81$\times$ the original design size and leads to a 143.72$\times$ simulation slowdown. 

Existing work does not fully solve the central challenge of deriving taint logic that is both lightweight enough for efficient simulation and precise enough to avoid excessive false positives.
For example, GLIFT~\cite{glift}, CellIFT~\cite{cellift}, and RTLIFT~\cite{rtlift} tackle this challenge by constructing precise taint logic at different abstraction levels. These approaches expose precision–cost tradeoffs by fixing the taint logic globally, i.e., across the entire circuit, at a chosen abstraction level.
Another line of work explicitly searches the precision--cost space of taint logic, 
but is either limited in terms of scalability~\cite{impreciseIFT2016,impreciseIFT2017} or
restricted to structural optimization that does not exploit the value-dependent semantics of taint propagation~\cite{hu2018property}.

A recent work, Compass~\cite{compass}, takes an important step forward by showing that taint-logic optimization can be cast as a counterexample-guided abstraction refinement (CEGAR)~\cite{cegar} process.
By refining taint logic from imprecise to more precise in response to spurious counterexamples (false positives), 
it demonstrates that taint-logic optimization can be property-driven and practical on processor-scale designs.
However, the refinement step in this approach remains \emph{heuristic-driven} and \emph{manually guided}: 
deciding where to refine relies on heuristic search, while deciding how to refine requires design knowledge and manual inspection. 
As a result, the approach is less systematic and harder to transfer across new designs and security properties.

We address these limitations with CEGAR-T, a \emph{fully automated} CEGAR-based framework for property-driven taint logic optimization.
The key insight is that taint-logic precision can be formalized semantically,
rather than treated as a fixed choice between an imprecise logic and a fully
precise one.
By defining precision with respect to NI semantics, CEGAR-T obtains
fine-grained control over the space of candidate taint logics for each hardware primitive block, enabling automated and incremental refinement.
Building on this structure, CEGAR-T introduces a novel Max-SAT~\cite{max-sat} formulation for design-level refinement, 
allowing the tool to select an
optimal set of local refinements to the primitive blocks sufficient to eliminate the current spurious counterexample.
This can be used iteratively in a CEGAR-T loop to eliminate successive spurious counterexamples.
Further, this can be combined with a SAT-based spuriouseness-check in our proposed two-stage CEGAR-T-based workflow to guarantee no false positives. 

We instantiate and implement CEGAR-T
on a word-level adaptation of CellIFT~\cite{cellift}.
We evaluate CEGAR-T in the setting of simulation-based IFT for the safe instruction set problem (SISP)~\cite{conjunct,veloCT}, a recently proposed processor-security problem for microarchitectural timing side-channel,
using open-source RISC-V cores Ibex~\cite{ibex}, Rocket~\cite{rocket}, and all four 
variants of BOOM~\cite{boom-v3}, 
which are widely used benchmarks for processor security evaluation.
Our results show that fully automated taint-logic refinement is scalable and yields significant simulation-based IFT speedup.

Our main contributions are:
\begin{itemize}

    \item We formalize taint-logic precision using the well-established non-interference semantics and derive a lattice-like refinement structure that characterizes \emph{how} 
    taint logic of a hardware primitive block can be refined to a more precise one.
    
    \item We formulate \emph{where} to refine as a Max-SAT optimization problem that selects a min-cost refinement set for a counterexample.

    \item We develop two complementary refinement algorithms for sequential logic designs with different optimality--scalability tradeoffs: 
    an unrolled algorithm that achieves per-counterexample min-cost refinements when tractable, and a more scalable single-frame alternative without sacrificing much optimality in practice.
    

    \item We implement CEGAR-T and evaluate it on simulation-based IFT of SISP. 
    Across the evaluated cores, 
    CEGAR-T reduces both instrumentation size and simulation time overhead, in geometric-mean, from 5.64$\times$ to 1.42$\times$ and from 34.65$\times$ to 1.79$\times$, respectively, without compromising the precision benefit of the  CellIFT baseline.
    
\end{itemize}

\section{Overview}
\label{sec:overview}



\subsection{Challenges in CEGAR-Based Optimization}
\label{sec:overview-challenges}


Counterexample-guided abstraction refinement (CEGAR)~\cite{cegar} was originally proposed as a formal verification paradigm for scaling model checking.
It reasons about an abstraction of the concrete system and refines it when the abstraction admits spurious counterexamples (cex). 
In CEGAR-based optimization for taint logic~\cite{compass}, the abstraction is the current taint logic.
It starts with imprecise taint logic, 
checks if a reported cex to the target information-flow property is spurious, 
and refines the logic to eliminate the spurious one.
The main challenge is refinement: eliminating a spurious cex while keeping taint logic lightweight requires deciding \emph{where} and \emph{how} to refine.

The \emph{where} question is intrinsically an optimization problem.
The objective is to eliminate the spurious cex with as few refinement sites, 
i.e., which hardware blocks to refine, 
as possible.
However, refinement decisions are globally coupled:
changing the taint logic of one block can alter which taints propagate to downstream blocks and, in sequential designs, across time-frames (i.e., clock-cycles) through state elements.
As a result, identifying a minimal set of refinement sites requires reasoning globally about taint propagation and its interaction with candidate refinements.

The \emph{how} question is equally nontrivial.
Even after a refinement site
is selected, one must decide how much precision to add to remove the spurious cex without unnecessarily resorting to the most precise and expensive taint logic.

In recent work that has used a CEGAR-based taint-logic optimization~\cite{compass},
these two challenges are handled through heuristic search for \emph{where} to refine and manual inspection for \emph{how} to refine.
While practical when guided by sufficient design knowledge and understanding of the target property, 
such refinement relies heavily on human effort, is less systematic, and harder to transfer across new designs and target properties.

\subsection{CEGAR-T: A Fully Automated Framework for Taint-Logic Optimization}
\label{sec:overview-cegar-t}

The proposed framework, CEGAR-T, tackles the \emph{how} challenge by first formalizing the well-established notion of precision using NI semantics. 
This serves as the foundation for defining a partial order over the space of taint logic with different precisions.
Such semantic characterization enables systematic strengthening of imprecise taint logic toward a more precise one.
Based on this notion, CEGAR-T derives a refinement lattice structure for
each block, capturing the space of taint logics between the initial
imprecise logic and the fully precise one.
As a result, \emph{how} to refine becomes a problem of navigating this
refinement lattice, enabling sound, incremental, and minimal changes to local taint logic over manually chosen updates or full replacement by the most precise and expensive one.

For the \emph{where} challenge, CEGAR-T formulates the refinement-site
selection as a constraint optimization problem using Max-SAT~\cite{max-sat}.
It encodes how candidate refinements alter taint propagation,
capturing the removal of the spurious
counterexample (cex) as constraints while minimizing the number of selected
refinement sites.
This enables global reasoning about interactions among
refinement choices, including cases where refining one block can suppress false
taints reaching downstream blocks and reduce the need for further
updates.

The introduced Max-SAT formulation captures how refinement choices affect taint propagation of the combinational logic within a clock cycle.
However, realistic RTL designs are sequential. 
A cex can
be a multi-cycle execution trace, where false taint may propagate across clock cycles through state elements.
To handle this temporal behaviour in practice, 
CEGAR-T provides two refinement algorithms that offer different optimality--scalability tradeoffs.
The first \emph{Unroll Approach} constructs the Max-SAT instance over the whole cex trace by \emph{unrolling}, which duplicates the combinational logic of the design for the number of cycles in the cex, allowing the solver to reason about the propagation across all cycles.
As a result, it computes a minimum-cost refinement for a given cex, but may incur high refinement time overhead due to unrolling as the cex gets longer.
The second \emph{Single-Frame Approach}, in contrast, avoids full unrolling by solving a sequence of simpler refinement problems within a single cycle. 
This sacrifices exact optimality for a given cex (with little empirical loss in terms of solution quality), but yields a significantly more scalable refinement procedure. 

\paragraph{Paper Roadmap}

After reviewing the required background in \Cref{sec:background}, \Cref{sec:spuriousness-check} presents a sound and complete procedure for checking whether a cex is spurious. 
\Cref{sec:taint-logic-precision-refinement} formalizes the notion of taint-logic precision and defines the corresponding refinement lattice. 
Building on this foundation, \Cref{sec:max-sat-optimum-refinement} develops the Max-SAT formulation for refinement site selection. 
Next, \Cref{sec:word-level-cellift} elaborates how the general CEGAR-T framework is instantiated on word-level CellIFT, \Cref{sec:workflow} describes a practical two-phase CEGAR-T-based simulation workflow, and \Cref{sec:evaluation} evaluates CEGAR-T in this setting.
\section{Background}
\label{sec:background}


\subsection{Notation}
We use $\bar x$ for the negation of a variable $x$,
$\neg \phi$ for negation of an expression $\phi$,
symbols ``$\cdot$'', ``$+$'', ``$\Rightarrow$'', and ``$\forall$'',  for logical 
AND, 
logical OR, implication, and universal quantification (forall), respectively.

\subsection{Information Flow Property}

An information flow property specifies that sources (secret inputs) must not affect the sink (attacker-observable output) in a design.
Intuitively, 
the sink should never change if we vary sources while holding other non-source inputs (public inputs) unchanged.
This is formalized by the non-interference (NI) property~\cite{non-interference}.
Consider a combinational logic design $y=f(X_p,X_s)$ where $y$ is the sink, $X_s$ the secret inputs, 
and $X_p$ the public ones.
The absence of information flow from $X_s$ to $y$ is captured by the NI condition below:
\begin{equation}
\forall X_p,\forall X_s,\forall X_s' \;.\;
f(X_p,X_s)=f(X_p,X_s'),
\label{eq:comb-noninterference}
\end{equation}
where $X_s'$ corresponds to a change to the secret inputs.
For a sequential design, 
we reason in terms of input traces and corresponding simulated executions.
An \emph{input trace} $\tau_K$ is a time-indexed sequence of input valuations over $k\in[0..K]$.
We write $X_p(k)$ and $X_s(k)$ for the public and secret parts of the input valuation at time $k$ in $\tau$.
Evaluating the design on $\tau_K$ induces a unique execution, 
which determines the valuations of all signals over time. 
We refer to this induced sequence of valuations as the \emph{full trace} of the design under $\tau_K$.
We denote the valuation of the sink in the full trace at time $k$ by
$y_{\text{sink}}(k)$.
NI requires that for all $K$ and for any two input traces $(\tau_K, \tau'_K)$ whose public inputs agree at every time step,
the sinks are identical at the last time step:

\begin{equation}
\begin{array}{rl}
     \forall K .\;\forall \tau_K,\tau'_K.\; \Big(\forall k.\; X_p(k)=X_p'(k)\Big)
\Rightarrow
( y_{\text{sink}}(K)=y_{\text{sink}}'(K) ),
\end{array}
\label{eq:seq-noninterference}
\end{equation}
where $y_{\text{sink}}(k)$ (resp., $y_{\text{sink}}'(k)$) is obtained by evaluating the design on $\tau_K$ (resp., $\tau'_K$).
If Eq.~\eqref{eq:seq-noninterference} is violated, 
then there is a $K$ and two input traces $\tau_K,\tau'_{K}$ such
that $X_p(k)=X_p'(k)$ for all $k \le K$, but $y_{\text{sink}}(K)\neq y_{\text{sink}}'(K)$.
The input traces serve as the cex to the property.

A standard way to check NI is via \emph{self-composition}~\cite{self-composition} following Eq.~\eqref{eq:comb-noninterference} and Eq.~\eqref{eq:seq-noninterference}. 
This approach builds a self-composed design consisting of two copies of the original design,
constrain their public inputs to be equal while leaving their secret inputs unconstrained, 
and assert equality of sinks across the two copies. 
For combinational logic, 
self-composition corresponds to Eq.~\eqref{eq:comb-noninterference},
whose violation can be checked via a SAT query over $(X_p,X_s,X_s')$. 
For sequential logic, 
self-composition yields a safety property~\cite{self-composition} in Eq.~\eqref{eq:seq-noninterference}, 
whose violation can be checked by a model checker.

\subsection{Hardware IFT with Taint Analysis}
\label{sec:background:taint-analysis}

Taint analysis is an alternative way to check the NI property that avoids the
two-copy self-composition. 
Instead, it instruments a \emph{single} copy of the design with additional taint bits and
taint logic, 
and then checks whether taint from the sources can reach the taint at the sink along an execution.

Taint analysis associates each signal $x$ with a taint bit $x^t$ indicating if there may be a flow from sources to $x$.
For a combinational block $y=f(X)$, 
a taint logic $f^t$ computes the output taint $y^t$ from the input values and input taints:
\begin{equation}
y^t = f^t(X,X^t).
\label{eq:taint-logic-bg}
\end{equation}
For a sequential design, 
sequential state elements such as D-flip flops (DFFs) propagate taint across cycles (e.g., $q^t(k{+}1)=d^t(k)$, where $q$ and $d$ are the DFF output and input, respectively), 
and taint is
propagated through combinational logic in each cycle with Eq.~\eqref{eq:taint-logic-bg}.
NI of both combinational and sequential cases is soundly approximated by checking if the sink could ever be tainted assuming sources are tainted ($X_s^t = \mathbf{1}$) and non-source inputs are untainted ($X_p^t = \mathbf{0}$).

Design-level taint logic is constructed \emph{compositionally}: 
each primitive block $f$ is assigned a taint logic $f^t$
(e.g., gates~\cite{glift}, RTL operators~\cite{rtlift}, or cells~\cite{cellift}), 
and composed according to the design connectivity. 
As a result, the sink taint is determined by the interaction of many local taint logics along the fanin cone of the sink and across cycles.

Taint analysis is designed to be a \emph{sound overapproximation} with respect to the NI semantics in
Eq.~\eqref{eq:comb-noninterference} and Eq.~\eqref{eq:seq-noninterference}.
In particular, if there exists an information-flow witness in the sense of NI, 
i.e., two executions that agree on all
public inputs but differ at the sink---then, 
when we label only the sources as tainted,
executing the design under any one of the two input traces must taint the sink in the taint logic.
However, the converse need not hold: 
the sink can become tainted even when no such paired execution exists,
leading to spurious information flows.



\paragraph{Scope of Refinement}

Imprecision arises primarily from two sources.
First, \emph{local imprecision} comes from using an imprecise taint logic for a primitive block $f$ that over-taints.
For example, 
the taint logic $a^t+b^t$ for a 2-input AND is imprecise since it ignores value-dependent blocking when a public input is $0$ (e.g., $(a^t, b^t)=(1,0)$ and $b=0$ implies $y=a\cdot b$ is independent of $a$, but $a^t+b^t$ still yields $y^t=1$). 
The precise taint logic considering the value-dependent effect is $a^t\cdot b^t + a^t \cdot b + a \cdot b^t$.
Second, \emph{correlation-based imprecision} arises even if each primitive is equipped with a precise taint logic in isolation. 
For instance, let $y=a\cdot b$ with $a=c\cdot d$ and $b=\bar c\cdot d$, and label only $c$ as a source ($c^t=1, d^t=0$), assuming precise taint logic for all the AND gates.
For the concrete valuation $(c, d)=(0, 1)$, 
we have $(a,b)=(0,1)$ and $(a^t,b^t)=(1,1)$, 
therefore yielding $y^t=1$. 
However, $y$ is constant zero for all $(c,d)$, 
so there is no actual flow from $c$ to $y$. 
The spurious taint occurs since local taint logic ignores the correlation between $a$ and $b$ induced by shared inputs ($b=1$ implies $a=0$).
These two sources of imprecision are discussed in the literature~\cite{glift,hu2011theoretical,compass}.
In this work,
we focus on eliminating local imprecision by refining per-block taint logic, 
and leave handling correlation-based imprecision as future work.
Based on our experience, 
the majority of the imprecision comes from local imprecision.


\subsection{Maximum Satisfiability (Max-SAT)}

(Weighted) maximum satisfiability~\cite{bacchus2021maximum} (Max-SAT) extends SAT with an optimization objective.
We use the following Max-SAT formulation (one of several equivalent ones~\cite{bacchus2021maximum}) as it aligns with our subsequent discussion and is supported by the solvers~\cite{Max-SAT1,Max-SAT2,Max-SAT3,Max-SAT4,Max-SAT5,z3-omt}.
Given a set of Boolean constraints $\Phi$ over variables $V$, Max-SAT searches for an assignment that satisfies $\Phi$ while
minimizing (or maximizing) a user-specified cost.
The cost is expressed as an objective over a subset of variables as a weighted sum.
We use the following notation for a Max-SAT query with constraint set $\Phi$ and objective $\min \sum_{i} w_i \cdot v_i$ with $v_i \in V$ a Boolean variable and $w_i\ge 0$ its weight (the unweighted case sets $w_i=1$):
\begin{equation}
\mathit{model} \;\gets\; \Call{MaxSAT}{\Phi,\ \min \sum_{i} w_i \cdot v_i}
\label{eq:maxsat-notation}
\end{equation}

There exists efficient Max-SAT solvers that scale to large instances~\cite{Max-SAT1,Max-SAT2,Max-SAT3,Max-SAT4,Max-SAT5}.
More generally, optimization modulo theories (OMT) solvers such as Z3's optimizing engine also support Max-SAT-style objectives with logical constraints over non-Boolean variables~\cite{z3-omt}.

\section{Spuriousness Check}
\label{sec:spuriousness-check}

Given a cex trace $\tau_K$ on the taint logic, 
i.e., an input trace under which the composed and possibly imprecise taint logic marks the sink as tainted,
we can check whether it corresponds to an actual violation of NI in two ways.
First, 
we can re-evaluate the same input trace $\tau_K$ under a precise (or more precise) reference taint logic (e.g., using precise taint logic for each primitive). 
If the sink is untainted with this reference taint logic,
then the cex must be spurious due to the overapproximation of taint logic.
We called this first approach \emph{precise taint simulation}.
Second, 
we can use the two-trace definition in Eq.~\eqref{eq:seq-noninterference}.
Concretely, 
we fix one side of the self-composed construction to the concrete execution induced by $\tau_K$, 
leave the other side symbolic but constrain public inputs to be equal at each time step, 
and ask whether there exists
another input trace $\tau'_K$ such that the sink differs at time $K$.
This reduces to a SAT query on the unrolled self-composed design; if unsatisfiable, the cex is spurious.

The SAT approach is complete for checking spuriousness but more expensive than precise taint simulation in general.
Precise taint simulation is sound and more efficient, but incomplete since, unlike the SAT approach, it does not account for correlation-based imprecision. 
In this work, we combine both approaches: 
given a cex,
we perform precise taint simulation first, 
and only use the SAT approach if the precise taint simulation produces a tainted sink.

A complete spuriousness check distinguishes real from spurious cex.
Note that when a spurious cex is detected, we can either let it trigger a refinement or just report it as a spurious cex.
This is explored further in Section~\ref{sec:workflow}.



\section{Precision and Refinement of Taint Logic}
\label{sec:taint-logic-precision-refinement}


We start by characterizing fine-grained precision for a single primary block of the design.
For this block, we propose a refinement lattice to enable minimum refinement that eliminates a given spurious cex.
 This is in contrast to prior precise taint construction~\cite{glift,rtlift,cellift}, which may lead to excessive refinement by forcing precision under all possible input values.

\subsection{Formalizing Per-assignment Precision}
\label{sec:precision}

We assume DFFs are the only sequential blocks in designs.
Since they simply propagate taint across clock cycles and cannot be refined, we focus on combinational blocks.

Let $X = (x_1,\ldots,x_n)$ be the inputs, 
$y = f(X)$ the output,
and $y^t = f^t(X, X^t)$ the taint logic of a block $f$.

A taint assignment $\tau^t \in \{0,1\}^n$ to $X^t$ partitions the inputs into the public inputs $X_p$ and the secret inputs $X_s$:
\begin{equation*}
\begin{array}{rcl}
X_p (\tau^t) & \triangleq & \{ x_i \mid x_i \in X \land x^t_i = 0 \text{ in $\tau^t$} \}, \\
X_s (\tau^t) & \triangleq & \{ x_i \mid x_i \in X \land x^t_i = 1 \text{ in $\tau^t$} \}.
\end{array}
\end{equation*}

As a reminder, this partition may change 
because it depends on how the hardware block is embedded within the larger design.
For the same block,
we may see different partitions in different cex or time frames.
We write the taint logic $f^t$ under the taint assignment $\tau^t$ as
\begin{equation*}
    f^t_{\tau^t}(X_p, X_s) \triangleq f^t(X, \tau^t).
\end{equation*}

When fixing a taint assignment $\tau^t$,
following the NI semantics in \cref{sec:background},
a taint logic $y^t=f^t_{\tau^t}(X_p,X_s)$ is \emph{precise} iff $y^t=1$ exactly when there is a flow from $X_s$ to $y$.
That is, 
there exists $X_s'$ such that $f(X_p,X_s)\neq f(X_p,X_s')$ with $X_p$ fixed,
as captured by the predicate $\hat{f}^t_{\tau^t}$:
\begin{equation}
\hat{f}^t_{\tau^t}(X_p, X_s) \triangleq \exists X_s' \;.\; f(X_p,X_s) \neq f(X_p,X_s').
\label{eq:delta}
\end{equation}
Given a taint logic $f^t$, we can now reason about its precision for \emph{a specific taint-assignment} $\tau^t$.
We say $f^t$ is \emph{precise under $\tau^t$} if $f^t_{\tau^t}$ is equivalent to $\hat{f}^t_{\tau^t}$.


The above concept of per-assignment precision guides us to make assignment-specific refinement in \Cref{sec:refinement} and to further derive our refinement lattice in \Cref{sec:refinement-lattice}.
Additionally, we give the definition of precise taint logic that is agnostic to taint assignments.
\begin{definition}[Precise taint logic]
\label{def:globally-precise}
A taint logic $f^t$ is \emph{precise} if it is precise under all taint
assignments:
\begin{equation}
\forall \tau^t \;.\; 
\text{$f^t$ is precise under $\tau^t$}
\label{eq:globally-precise}
\end{equation}
\end{definition}

\begin{example}
\label{ex:and2}
Let $X=(a,b)$ and $y=f(a, b)=a \cdot b$.
We consider two candidate taint logics $f_i^t$ with index $i \in \{0,1\}$ denoting different taint logics for the same function $f$:

\begin{equation*}
\begin{array}{rclcrcl}
f^t_0(a, b, a^t, b^t) &=& a^t + b^t, && \\
f^t_1(a, b, a^t, b^t) &=& a^t + (a \cdot b^t).
\end{array}
\end{equation*}

Case (1): $a$ public, $b$ secret.
Let $\tau^t=(0,1)$, so $X_p=\{a\}$ and $X_s=\{b\}$.
We compute $\hat f^t_{(0, 1)}$ by instantiating Eq.~\eqref{eq:delta}:
\begin{equation*}
\hat f^t_{(0,1)} =  \exists b' \;.\; (a \cdot b) \neq (a \cdot b') = a.
\end{equation*}
Under $\tau^t=(0,1)$, the two taint logics simplify to: 
\begin{equation*}
\begin{array}{rclcrcl}
f^t_0(a, b, 0, 1)  &=& 1, &&
f^t_1(a, b, 0, 1)  &=& a.
\end{array}
\end{equation*}
Therefore $f^t_1$ is precise under $\tau^t = (0,1)$ (it matches $\hat f^t_{(0,1)} = a$),
whereas $f^t_0$ is imprecise.

Case (2): $a$ secret, $b$ public.
Let $\tau^t=(1, 0)$, so $X_p=\{b\}$ and $X_s=\{a\}$.
Here, both taint logics reduce to $1$ under $\tau^t$:
\begin{equation*}
\begin{array}{rclcrcl}
f^t_1 (a, b, 1, 0 ) &=& 1, &&
f^t_2 (a, b, 1, 0 )  &=& 1,
\end{array}
\end{equation*}
, which are in-equivalent to $\hat f^t_{(1, 0)} = b$, 
and thus imprecise under $\tau^t=(1,0)$. 

\paragraph{Constructing Precise Taint Logic.}
In fact, 
with \Cref{def:globally-precise}, 
we are able to construct the precise taint logic.
Let the minterms over $(a^t,b^t)$ be:
\begin{equation*}
\begin{array}{rclcrcl}
m_{00} &\triangleq& \bar{a}^t \cdot \bar{b}^t, &&
m_{01} &\triangleq& \bar{a}^t \cdot b^t, \\[2pt]
m_{10} &\triangleq& a^t \cdot  \bar{b}^t, &&
m_{11} &\triangleq& a^t \cdot b^t. 
\end{array}
\end{equation*}

A precise taint logic can be constructed by a minterm expansion over $\tau^t=(a^t,b^t)$.
For each taint assignment, we gate the corresponding per-assignment 
precision predicate $\hat f^t_{(i,j)}$
with the matching minterm $m_{ij}$ of $(a^t,b^t)$ and take the disjunction of the results, as prescribed by \Cref{def:globally-precise}.
For the 2-input AND cell, we have $\hat f^t_{(0,0)}=0$, $\hat f^t_{(0,1)}=a$,
$\hat f^t_{(1,0)}=b$, and $\hat f^t_{(1,1)}=1$. 
This yields:
\begin{equation}
\begin{array}{rcl}
\tilde{f}^t
&=& (m_{00} \cdot \hat f^t_{(0,0)}) + (m_{01} \cdot \hat f^t_{(0,1)})\ + \\[2pt]
&& (m_{10} \cdot \hat f^t_{(1,0)}) + (m_{11} \cdot \hat f^t_{(1,1)}) \\[2pt]
&=& (\bar{a}^t \cdot b^t \cdot a) + (a^t \cdot \bar{b}^t \cdot b) + (a^t \cdot b^t),
\end{array}
\label{eq:and2-precise}
\end{equation}
which is equivalent to the precise taint logic in the literature~\cite{glift} for 2-input AND $(a^t \cdot b) + (a \cdot b^t) + (a^t \cdot b^t)$.

We remark that such a precise taint logic construction is different from the canonical $m$-replica architecture in CellIFT~\cite{cellift} but yields the functionally equivalent precise taint logic.
In subsequent discussion, given a function $f$,
we assume having access to the precise taint logic $\tilde{f}^t$.


\end{example}

\subsection{Counterexample-Guided Precision Refinement}
\label{sec:refinement}

We next explain how to refine a taint logic $f^t$ to remove a spurious cex in which $f^t$
\emph{over-taints}.

Let's consider a motivating example of a 3-input AND cell 
$y=a\cdot b \cdot c$
with the most imprecise taint logic
\begin{equation*}
    f^t_{0}=a^t + b^t + c^t.
\end{equation*}
Suppose we observe a cex assignment with $\tau^t = (1, 0, 0)$ and  $\tau=(1, 0, 1)$, under which taint should \emph{not}
propagate to the output, but the current taint logic
nevertheless 
sets $y^t=1$.
We describe three versions of refinement to demonstrate the benefit of making assignment-specific refinements.

First, a naive refinement would block exactly this single valuation by conjoining a constraint that excludes it:
\begin{equation*}
f^t_{\text{naive}}
=(a^t+b^t+c^t)\cdot \neg(a^t\cdot \bar{b}^t\cdot \bar{c}^t\cdot a\cdot \bar{b}\cdot c).
\end{equation*}
However, this refinement is overly specific: it eliminates only the particular cex and may still admit many
other assignments that share the same taint assignment $\tau^t=(1,0,0)$ and are equally spurious. For example,
under the same $\tau^t=(1,0,0)$,
$(a,b,c)=(0,0,1)$ and $(1,0,0)$ should also yield $y^t=0$ because $b=0$ or $c=0$
blocks the secret $a$, yet $f^t_0$ over-taints these cases as well.

Second, at the other extreme, we could replace $f^t_0$ with $\tilde{f}^t$,
thereby eliminating \emph{all} spurious over-tainting for this cell. 
While effective, this can be
unnecessarily expensive when $\tilde{f}^t$ contains terms that are irrelevant under the
design's reachable behaviors, e.g., the don't-care values in the design.

Lastly, we seek a refinement that generalizes from the cex to capture the \emph{blocking condition}
over the public inputs,
which
characterizes the public input values that block the taints for $f^t$ and the given $\tau^t$.

Under $\tau^t=(1,0,0)$, the secret is $X_s = \{a\}$ and the public inputs are $X_p=\{b,c\}$.
For $y=a\cdot b\cdot c$, the secret can influence $y$ iff $b\cdot c=1$; equivalently, $(\bar{b}+\bar{c})$ blocks
taint propagation. 
Thus, a \emph{just-enough refinement}, that lies in between the two ends above, strengthens $f^t_0$ by ruling out taint
propagation whenever $(\bar{b}^t\cdot \bar{c}^t\cdot (\bar{b}+\bar{c}))$ holds:
\begin{equation*}
\begin{array}{rcl}
f^t_{\text{ref}}
&=& (a^t+b^t+c^t)\cdot \neg (\bar{b}^t\cdot \bar{c}^t\cdot (\bar{b}+\bar{c})).
\end{array}
\label{eq:just-enough-and3}
\end{equation*}
As a result, 
this refinement eliminates a set of generalized cex, 
the original cex and a family of cex that share the same taint assignment $\tau^t=(1,0,0)$, using a simpler taint logic compared with $\tilde{f}^t$.
This potentially avoids generating cex from the same family,
thus reducing the number of iterations in the CEGAR-T loop.


\paragraph{Refine for Arbitrary Blocks}
We now formalize the above intuition for an arbitrary combinational logic $f$ and its taint logic $f^t$.
Let a cex provide a taint assignment $\tau^t$ and concrete valuations
$(\tau_p,\tau_s)$ for $(X_p,X_s)$.
We compute a blocking condition $R_{\tau^t}$ over public inputs under $\tau^t$:
\begin{equation}
\begin{array}{rcl}
R_{\tau^t}(X_p) &\triangleq& \forall X_s, X'_s \;.\; f(X_p,X_s)=f(X_p,X'_s).
\end{array}
\label{eq:blocking-R}
\end{equation}
Intuitively, $R_{\tau^t}(X_p)=1$ means that for the given $X_p$, varying secret inputs cannot change the output, so the secret does not propagate.
Using $R_{\tau^t}$, we refine $f^t$ by ruling out taint propagation whenever all public inputs are untainted and the
blocking condition holds. 
Writing $(X_p^t=\mathbf{0}) \triangleq \prod_{x_i\in X_p}\bar{x}^t_i$, the refined taint logic is:
\begin{equation}
\begin{array}{rcl}
f^t_{\text{ref}}
&=& f^t \cdot \neg\big((X_p^t=\mathbf{0})\cdot R_{\tau^t}(X_p)\big).
\end{array}
\label{eq:cex-refine}
\end{equation}
Note that the blocking condition $R_{\tau^t}$ depends on $\tau^t$, not on the concrete assignment for the public/secret inputs $\tau_p/\tau_s$, 
i.e., 
we refine the taint logic w.r.t. the taint assignment $\tau^t$, 
which captures a family of cex of the same $\tau^t$.

\subsection{Refinement Lattice}
\label{sec:refinement-lattice}


We now formalize the space of candidate taint logics by viewing each application of Eq~\eqref{eq:cex-refine} as a refinement step to a \emph{more precise} taint logic.

\paragraph{Root node (most imprecise taint logic).}
In theory, the constant taint logic $y_t = 1$ is the most imprecise one but it is not practically useful. 
Following prior work~\cite{glift,impreciseIFT2016,impreciseIFT2017}, we use the disjunction of input taints as the root node.
\begin{equation}
f^t_0(X,X^t) \triangleq \sum_{i=1}^n x_i^t.
\label{eq:ft-init}
\end{equation}

\paragraph{Refinement operator.}
Let $\tau^t \in \{0,1\}^n$ be a taint assignment inducing the partition $(X_p,X_s)$.
We define the refinement operator by directly rewriting Eq.~\eqref{eq:cex-refine} for an arbitrary current taint logic $f^t$:
\begin{equation}
\begin{array}{rcl}
\mathsf{Refine}(f^t,\tau^t)
&\triangleq& f^t \cdot \neg\big((X_p^t=\mathbf{0})\cdot R_{\tau^t}(X_p)\big).
\end{array}
\label{eq:refine-op}
\end{equation}
We only apply $\mathsf{Refine}(\cdot,\tau^t)$ when $f^t$ is imprecise under $\tau^t$.

\begin{definition}[Refinement lattice]
\label{def:refinement-lattice}
Given a function $f$, let $\mathcal{T}\triangleq \{0,1\}^n$ be the set of all possible taint assignments.
The \emph{refinement lattice} for $f$ is the directed graph $\mathcal{L}=(V,E)$:
\begin{itemize}
\item \textbf{Nodes.} $V$ is the smallest set of taint logics such that:
(i) $f^t_0\in V$, and
(ii) for any $f^t\in V$ and any $\tau^t\in \mathcal{T}$, if $f^t$ is imprecise under $\tau^t$, then
$\mathsf{Refine}(f^t,\tau^t)\in V$.
\item \textbf{Edges.} For any $f^t\in V$ and $\tau^t\in\mathcal{T}$ such that $f^t$ is imprecise under $\tau^t$,
we add a labeled edge:
\begin{equation*}
f^t \xrightarrow{\;\tau^t\;} \mathsf{Refine}(f^t,\tau^t).
\end{equation*}
\end{itemize}
\end{definition}

Each node in $\mathcal{L}$ is a candidate taint logic reachable from $f^t_0$ by repeatedly applying
$\mathsf{Refine}(\cdot,\tau^t)$, and thus represents a different precision point in our search space.
Moreover, the precise taint logic $\tilde f^t$ is the sink node in $\mathcal{L}$, since no refinement is applicable once a taint logic is precise under every taint assignment.

Prior work~\cite{impreciseIFT2016} constructs a family of imprecise taint logics by introducing \emph{don’t-care} conditions over  $X$, 
starting from a precise taint logic and simplifying it via offline precomputation.
In contrast, our refinement lattice is built around \emph{taint assignments} over $X^t$: 
each edge corresponds to strengthening $f^t$ by the blocking condition,
enabling implicit, on-demand traversal starting from a cheap root logic.
This yields a finer-grained space, e.g., for a 3-input AND gate, their construction yields 8 variants, 
whereas our lattice contains 64 nodes and subsumes their variants.

\paragraph{Key properties.}
From Eq.~\eqref{eq:refine-op}, each refinement step strengthens the current taint logic by conjoining a constraint.
As a result, refinement steps commute: for any $f^t$ and $\tau_1^t,\tau_2^t\in\mathcal{T}$,
\begin{equation*}
\mathsf{Refine}(\mathsf{Refine}(f^t,\tau_1^t),\tau_2^t)
\;\equiv\;
\mathsf{Refine}(\mathsf{Refine}(f^t,\tau_2^t),\tau_1^t),
\end{equation*}
This property ensures that when a block requires multiple refinements, 
which is needed later in \cref{sec:max-sat-optimum-refinement} for multi-cycle cex,
the applications of refinements are order-independent.

For a set of taint assignments $T\subseteq \mathcal{T}$ from multiple cex,
we define the \emph{batch-refined} taint logic as:
\begin{equation}
\begin{array}{rcl}
\mathsf{Refine}(f^t,T)
&\triangleq&
f^t \cdot \prod_{\tau^t\in T} G_{\tau^t},
\end{array}
\label{eq:batch-refine}
\end{equation}
where $G_{\tau^t}$ abbreviates the guard conjoined in Eq.~\eqref{eq:refine-op}.
By commutativity of conjunction, Eq.~\eqref{eq:batch-refine} is equivalent to applying
$\mathsf{Refine}(\cdot,\tau^t)$ sequentially for all $\tau^t\in T$ in any order.
In sequential designs, multiple cex for the same $f_t$ may arise at different cycles of one execution trace.
Batch refinement allows us to refine multiple cex at once.

\begin{example}
\label{ex:and2_lattice}
\textbf{\begin{figure}[t]
    \centering
    \includegraphics[width=.6\columnwidth]{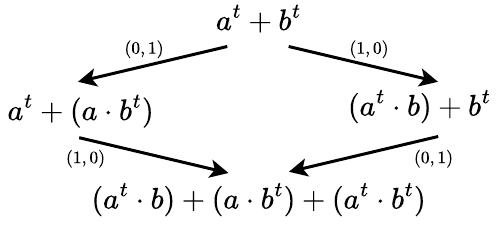}
    \caption{The refinement lattice for $y = a \cdot b$.}
    \label{fig:and2-lattice}
\end{figure}}
Consider again the 2-input AND with root taint logic
$f^t_0=a^t+b^t$.
There are four taint assignments $\mathcal{T}=\{(0,0),(0,1),(1,0),(1,1)\}$.
We have that $f^t_0$ is precise under $(0,0)$ and $(1,1)$, but imprecise under
$(0,1)$ and $(1,0)$.
By \Cref{def:refinement-lattice}, there are two outgoing edges from $f^t_0$ in the lattice 
for the two imprecise
taint assignments:
\begin{equation*}
\begin{array}{rcl}
\mathsf{Refine}(f^t_0,(0,1)) &\equiv& f^t_1 \triangleq a^t + (a\cdot b^t),\\[2pt]
\mathsf{Refine}(f^t_0,(1,0)) &\equiv& f^t_2 \triangleq b^t + (a^t \cdot b).
\end{array}
\end{equation*}
Thus, the lattice contains edges $f^t_0 \xrightarrow{(0,1)} f^t_1$ and
$f^t_0 \xrightarrow{(1,0)} f^t_2$.
Refining once more from either intermediate node yields the precise taint logic:
\begin{equation*}
\begin{array}{rcl}
\mathsf{Refine}(f^t_1,(1,0))
&\equiv&
\mathsf{Refine}(f^t_2,(0,1)) \\
&\;\equiv\;&
\tilde f^t \;=\; (a^t\cdot b) + (a\cdot b^t) + (a^t\cdot b^t).
\end{array}
\end{equation*}
The visualization of the lattice is shown in~\Cref{fig:and2-lattice}.

\end{example}


\section{Max-SAT for Optimum Refinement}
\label{sec:max-sat-optimum-refinement}

In a design, sink taint is determined by the interaction of many local taint logics.
\cref{sec:taint-logic-precision-refinement} defined the search space of candidate taint logic as a refinement lattice spanned by the $\mathsf{Refine}$ operator.
This allows us to perform cex-guided refinement for a \emph{single block}.
This section addresses the remaining \emph{refinement selection} problem: given a spurious cex with a falsely tainted sink,
which blocks in the design must be refined to eliminate the sink taint while minimizing the added instrumentation cost?

We consider a cex in the current taint logic to the sink taint, 
i.e., a concrete execution trace
under which the composed taint logic marks the sink as tainted ($y^t_{\text{sink}}=1$).
Constructing a \emph{precise reference} model (e.g. CellIFT~\cite{cellift}) obtained by using the precise taint logic
$\tilde{f}^t$ for each block, our goal is to refine a selected set of combinational blocks so that re-evaluating the same trace leaves the sink untainted.

Among all refinements that eliminate the cex, we aim to minimally change the taint logic of the design.
Concretely, we seek a refinement that \emph{minimizes the number of blocks whose taint logic is refined}.
While this metric does not exactly translate to the final complexity of the taint logic, 
it is a reasonable proxy in practice.
At a high level, 
this optimization
can be formulated as a Max-SAT problem.
We associate a variable $r_f$ with each combinational block $f$, where $r_f=1$ iff we refine the taint logic $f^t$ to a more precise one.
Intuitively, we want to search for a set of blocks to refine (i.e., set some $r_f$ to $1$) such that the resulting composed taint logic
makes the sink untainted on the cex ($y^t_{\text{sink}}=0$), 
while setting as few $r_f$ to $1$ as possible.
This corresponds exactly to the Max-SAT problem for finding a satisfiable solution over $r_f$ variables, 
while minimizing the number of $1$'s in the solution.


\subsection{Identifying Refinement Candidate Set}

Not every taint observed in the current taint logic can (or should) be eliminated by refinement.
In fact, 
any taint seen by the precise reference model must be preserved. 
Refinement should only address blocks whose output is tainted in the current model but untainted in the reference model, i.e., the output is \emph{falsely tainted}. 
Therefore, 
before formulating the Max-SAT instance, 
we identify combinational blocks with false taints under the current taint logic. 
We refer to this set of blocks as a \textit{refinement candidate set}.

Given the cex trace, 
we perform two taint simulations on the unrolled design:
(i) the current taint simulation using the current block taint logic $f^t$, 
and
(ii) the reference taint simulation using the precise reference model.
For each combinational block $f$ and timeframe $k$, 
let $Y^t(f,k)$ and $\tilde Y^t(f,k)$ denote the
taint value of $f$'s output under the two simulations, respectively.
We add $f$ to the refinement candidate set $\mathcal{C}$ if it is falsely tainted in at least one timeframe:
\begin{equation}
\mathcal{C} \triangleq \{\, f \mid \exists k \;.\; Y^t(f,k)=1 \text{ and } \tilde Y^t(f,k)=0 \,\}.
\label{eq:candidate-set}
\end{equation}
Only blocks in $\mathcal{C}$ are eligible to be refined for this cex.
Conceptually, $\mathcal{C}$ excludes blocks whose taint outputs already agree with the reference model, so
refining them would not help eliminate any over-tainting observed on this cex.

\begin{figure}[t]
    \centering
    \includegraphics[width=.8\columnwidth]{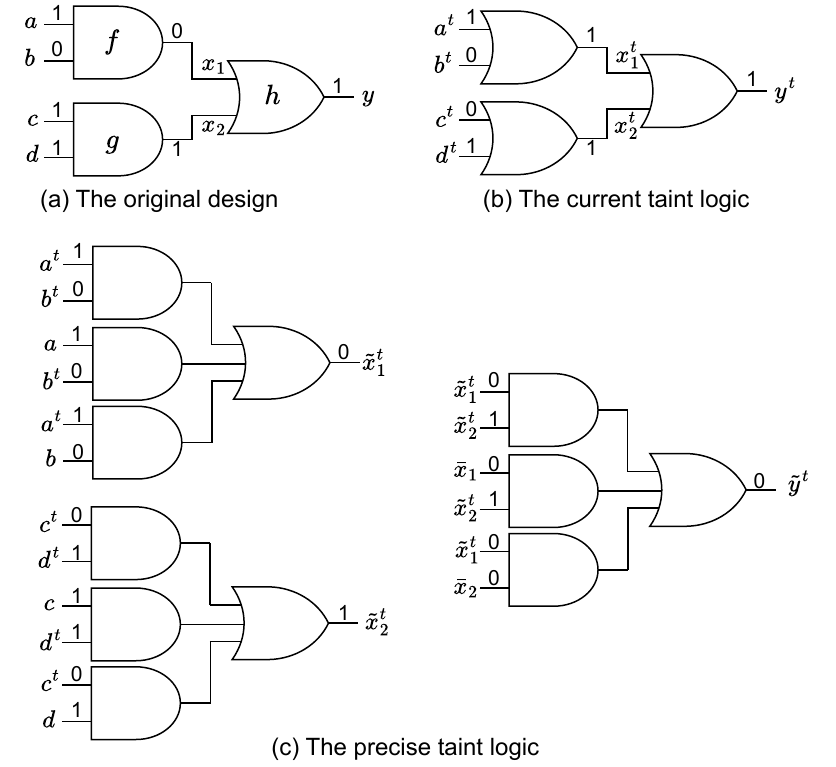}
    \caption{The simulation results of the cex $(a,b,c,d)=(1, 0, 1, 1)$ and $(a^t, b^t, c^t, d^t) = (1, 0, 0, 1)$ on (a) original design, (b) current taint logic, and (c) precise taint logic.}
    \label{fig:two-simulation}
\end{figure}

\begin{example}
    \label{ex:refinement-candidate}
    Consider a design with inputs $(a, b, c, d)$ and an output $y$.
    Assume that the current taint logic uses the most imprecise one for all blocks, 
    and that we encounter a cex with $\tau = (1, 0, 1, 1)$ and $\tau^t = (1, 0, 0, 1)$, 
    as shown in \Cref{fig:two-simulation}~(a) and \Cref{fig:two-simulation}~(b). 
    By simulating the precise taint logic with $\tau^t$ and $\tau$, 
    we acquire the reference taint values $(\tilde x_1^t, \tilde x_2^t, \tilde y^t) = (0, 1, 0)$ as shown in \Cref{fig:two-simulation}~(c).
    Since the output taints in the current taint logic are inconsistent with the ones in the precise taint logic for all blocks, 
    the refinement candidate set is $\mathcal{C} = \{f, h\}$.
    
\end{example}


\subsection{Circuit Construction for Refinement Candidates}
\label{sec:candidate-mux}


For each refinement candidate block $f\in\mathcal{C}$, 
we use the variable $r_f$ indicating whether we
apply refinement to $f^t$ for the current cex.
A first attempt is to gate away the false taint by a mux: if $r_f=1$ we force the block's output taint to $0$, otherwise we
keep the original output taint computed by the current taint logic $f^t$.
However, this construction is not sound: it may suppress output taint even when the current taint logic already matches the precise reference model.

The key subtlety is that the input-taint assignment seen by $f^t$ depends on the taints of its upstream fanin logic.
Since some fanin taints may be falsely tainted and eliminated by refining upstream blocks, 
the input taints $X^t$ arriving at $f^t$ can change depending on upstream refinements.
Thus, 
whether the false taint at $f^t$ \emph{can} be eliminated by refinement must be based on the input-taint $\tau^t$ assignment
reaching $f^t$, referred to as the \emph{incoming input-taint assignment}, 
together with the input assignment $\tau$ from the cex.

To capture this dependency, we construct an enabling condition $S_f(X^t)$ over the input taint bits $X^t$.
Let $\tau$ be the assignment to $X$ given by the cex.
We define $S_f$ to describe when $f^t$ is falsely tainted on the cex under the incoming input-taint
assignment:
\begin{equation}
S_f(X^t) \triangleq \big( f^t(\tau, X^t) \neq \tilde f^t(\tau, X^t) \big).
\label{eq:sf-cex}
\end{equation}
Intuitively, $S_f(X^t)=1$ means that, 
under $\tau$ and the seen $X^t$, 
the output taint produced by $f^t$ disagrees with the precise reference model, 
and hence is imprecise and eligible to be fixed by refinement.
When $S_f(X^t)=0$, the current taint logic already matches the reference model, 
and we should not
allow refinement to change the output taint.

We then guard refinement by $S_f$ so that the refinement is enabled only when (i) the solver chooses $r_f=1$ and (ii) the incoming input-taint assignment makes $f^t$ falsely tainted.
Concretely, letting $y^t$ be the output taint of $f$, 
we enforce
\begin{equation}
y^t \;=\; (r_f \cdot S_f(X^t)) \ternop 0 \terncol f^t(X, X^t).
\label{eq:candidate-mux-guarded}
\end{equation}
This ensures that a block's output taint can be suppressed only in cases where it is falsely tainted on the cex under the
incoming input-taint assignment reaching the block.

\begin{figure}[t]
    \centering
    \includegraphics[width = .8\columnwidth]{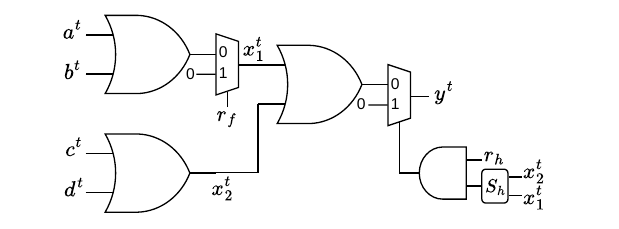}
    \caption{The circuit construction of \Cref{ex:mux-construction} for the cex used in \Cref{ex:refinement-candidate}. The $S_h$ block implements $\bar x_1^t \cdot x_2^t + x_1^t \cdot \bar x_2^t$.}
    \label{fig:mux-construction}
\end{figure}

\begin{example}
    \label{ex:mux-construction}
    Following \Cref{ex:refinement-candidate}, the cex is $\tau=(1,0,1,1)$, $\tau^t=(1,0,0,1)$ (\Cref{fig:two-simulation}),
and the refinement candidate set is $\mathcal{C}=\{f,h\}$.
The Max-SAT circuit construction is shown in \Cref{fig:mux-construction}.
For $f$, since there is no other upstream logic that could affect its input taints, its guard is effectively constant one and simplified away.
For $h$, following Eq.~\eqref{eq:candidate-mux-guarded}, 
we derive $S_h(x_1^t, x_2^t) = \bar x_1^t \cdot x_2^t + x_1^t \cdot \bar x_2^t$ and use $S_h$ to guard the fix variable $r_h$.
The Max-SAT problem of this circuit would be finding an assignment to $(r_f, r_h)$ such that $y^t$ becomes $0$, while minimizing the number of ones in the solution.
There is one optimum solution $(r_f, r_h)=(1, 1)$.

A Max-SAT solution not only selects which blocks to fix, 
but also implicitly identifies the
\emph{refinement step} in the lattice, 
by determining the input-taint assignments under which the refinement is applied for each refined block.
For example, consider the solution $(r_f, r_h) = (1, 1)$.
For $f^t$, since it sees the taint assignment $(c^t, d^t) = (1, 0)$, 
the refined taint logic would be $\mathsf{Refine}(f^t,(1,0)) = (c^t \cdot d) + d^t$.
For $h^t$, since its upstream taint $x_2^t$ from $g^t$ is fixed, it sees the input-taint assignment $(x_1^t, x_2^t) = (0, 1)$.
The refined taint logic is $\mathsf{Refine}(h^t,(0,1)) = x_1^t + (\bar x_1 \cdot x_2^t)$.


\end{example}



\subsection{Overall Algorithm}
\label{sec:overall-alg}

With (i) the refinement candidate set,
(ii) the guarded mux construction for candidates,
and (iii) the refinement operator,
we now present the end-to-end procedure\soohyuktext{s} for eliminating a spurious cex
on a general sequential design. 

\subsubsection{Unroll Approach}
\label{sec:unroll-approach}
The pseudocode of the unroll approach is shown in \Cref{alg:unroll-maxsat}.
Given a cex trace $\tau$ of length $K$, we run two simulations on the unrolled design: 
one using the current taint logic ${f^t}$ and one using the precise reference logic ${\tilde f^t}$ (lines 1--2). 
From the mismatch between the two runs, we identify a refinement candidate set $\mathcal{C}$ consisting of combinational blocks that are falsely tainted in at least one timeframe (line 3). 
For each candidate block $f\in\mathcal{C}$, we introduce the variable $r_f$ (line 4)
to indicate if $f$ requires refinement to eliminate spurious taint on this trace.
We then construct an unrolled Max-SAT instance $D'$ over frames $k\in[0..K]$ (lines 5–-11). 
For each combinational block instance in frame $k$, we add the corresponding taint constraint in topological order. 
Non-candidates are encoded using their current taint logic. 
For candidates, 
we use the guarded mux construction (Eq.~\eqref{eq:candidate-mux-guarded}) via the helper function \Call{AddTaintConstraint}{} shown in \Cref{alg:add-taint-constraint} for the circuit construction.
Sequential elements are handled with the identity constraint for each DFF 
(lines 9–-10).
Finally, we solve one Max-SAT problem (line 12) that enforces the sink to be untainted at the last frame (line 11), 
while minimizing $\sum_{r_f\in\mathcal{R}} r_f$.
The resulting model determines which blocks to refine, 
and implicitly the frames those refinements were actually enabled.
For each selected block $f$ with $\mathit{model}(r_f)=1$ in the MAX-SAT solution, 
we collect the set $\Gamma_f$ of input-taint assignments $X^t(f,k)$ for frames where the guard was active (lines 15--17), 
and then update the taint logic by applying batch refinement (line 18).

\begin{algorithm}[t]
\caption{A helper function used by \Cref{alg:unroll-maxsat}  assuming we have access to the fix variable $r_f$ for $f \in \mathcal{C}$.}
\label{alg:add-taint-constraint}
\begin{algorithmic}[1]
\Procedure{\textsc{AddTaintConstraint}}{$D', f, k, \tau, \mathcal{C}$}
    \If{$f \notin \mathcal{C}$}
        \State \begin{tabular}[t]{@{}l@{}}
        Add constraint to $D'$: \\ 
        $y^t(f,k) = f^t(\tau(f,k), X^t(f,k))$
        \end{tabular}
    \Else
        \State $s_{f,k} \gets S_{f,k}(X^t(f,k))$ \Comment{instantiate Eq.~\eqref{eq:sf-cex}}
        \State \begin{tabular}[t]{@{}l@{}}
        Add constraint to $D'$: \\
        $y^t(f,k) \;=\; (r_f \cdot s_{f,k}) \ternop 0 \terncol f^t(\tau(f,k),X^t(f,k))$
        \end{tabular}
    \EndIf
\EndProcedure
\end{algorithmic}
\end{algorithm}

\begin{algorithm}[t]
\caption{Unroll-based Max-SAT refinement.}
\label{alg:unroll-maxsat}
\begin{algorithmic}[1]
\Require Design $D=(D_{\text{comb}},D_{\text{dff}})$ and its current taint logic $\{f^t\}$; precise taint logic $\{\tilde f^t\}$;
spurious cex $\tau$ of length $K$; sink $y_{\text{sink}}$

\State $Y^t \gets \Call{Sim}{D,\{f^t\},\tau,K}$ \Comment{full trace of $(D, \{f^t\})$ on $\tau$}
\State $\tilde Y^t \gets \Call{Sim}{D,\{\tilde f^t\},\tau,K}$
\Comment{full trace of $(D, \{ \tilde f^t \} )$ on $\tau$}
\State $\mathcal{C} \gets \{\, f\in D_{\text{comb}} \mid \exists k.\ Y^t(f,k)=1 \text{ and } \tilde Y^t(f,k)=0 \,\}$
\State $\mathcal{R} \gets \{\, r_f \mid f\in\mathcal{C} \,\}$

\State $D' \gets \emptyset$
\For{$k \gets 0$ to $K$}
    \ForAll{$f \in D_{\text{comb}}$ in topological order}
        \State \Call{AddTaintConstraint}{$D'$, $f$, $k$, $\tau$, $\mathcal{C}$} 
    \EndFor
    \ForAll{$q \in D_{\text{dff}}$ with input $d$}
        \State Add constraint to $D'$:\ $q^t(k{+}1) = d^t(k)$
    \EndFor
\EndFor
\State Add constraint to $D'$:\ $y^t_{\text{sink}}(K)=0$
\State $\mathit{model} \gets \Call{MaxSAT}{D',\ \min \sum_{r_f\in\mathcal{R}} r_f}$

\ForAll{$f\in\mathcal{C}$ with $\mathit{model}(r_f)=1$}
    \State $\Gamma_f \gets \emptyset$
    \For{$k \gets 0$ to $K$}
        \If{$\mathit{model}(S_{f,k}(X^t(f,k)))=1$}
            \State $\Gamma_f \gets \Gamma_f \cup \{\mathit{model}(X^t(f,k))\}$
        \EndIf
    \EndFor
    \State $f^t \gets \mathsf{Refine}(f^t,\Gamma_f)$ \Comment{batch refinement, Eq.~\eqref{eq:batch-refine}}
\EndFor
\end{algorithmic}
\end{algorithm}

\begin{algorithm}[t]
\caption{Single-frame Max-SAT refinement.}
\label{alg:single-frame-maxsat}
\begin{algorithmic}[1]
\Require Design $D=(D_{\text{comb}},D_{\text{dff}})$ and its current taint logic $\{f^t\}$; precise taint logic $\{\tilde f^t\}$;
spurious cex trace $\tau$ of length $K$; sink signal $y_{\text{sink}}$

\State $Y^t \gets \Call{Sim}{D,\{f^t\},\tau,K}$ 
\Comment{full trace of $(D, \{f^t\})$ on $\tau$}
\State $\tilde Y^t \gets \Call{Sim}{D,\{\tilde f^t\},\tau,K}$
\Comment{full trace of $(D, \{ \tilde f^t \})$ on $\tau$}

\For{$k \gets 0$ to $K$}
    \If{$y^t_\text{sink}=0$ in $Y^t (K)$}
        \State \textbf{break}
    \EndIf

    \State $\mathcal{C}_k \gets \{\, f\in D_{\text{comb}} \mid Y^t(f,k)=1 \text{ and } \tilde Y^t(f,k)=0 \,\}$
    \State $\mathcal{R}_k \gets \{\, r_f \mid f\in\mathcal{C}_k \,\}$

    \If{$k < K$}
        \State $\mathcal{F} \gets \{\, q\in D_{\text{dff}} \mid Y^t(q,k{+}1)=1 \text{ and } \tilde Y^t(q,k{+}1)=0 \,\}$
        \State $\Phi_k \gets (\mathcal{F} = \mathbf{0})$ 
    \Else
        \State $\Phi_k \gets (y^t_{\text{sink}}(K)=0)$ 
    \EndIf

    \State $D'_k \gets \emptyset$
    \ForAll{$f \in D_{\text{comb}}$ in topological order}
        \State \Call{AddTaintConstraint}{$D'_k$, $f$, $k$, $\tau$, $\mathcal{C}_k$}
    \EndFor
    \ForAll{$q \in D_{\text{dff}}$ with data input $d$}
        \State Add constraint to $D'_k$:\ $q^t(k{+}1) = d^t(k)$
    \EndFor
    \State Add constraint to $D'_k$:\ $\Phi_k$
    \State $\mathit{model} \gets \Call{MaxSAT}{D'_k,\ \min \sum_{r_f\in\mathcal{R}_k} r_f}$

    \ForAll{$f\in\mathcal{C}_k$ with $\mathit{model}(r_f)=1$}
        \If{$\mathit{model}(S_{f,k}(X^t(f,k)))=1$}
            \State $f^t \gets \mathsf{Refine}(f^t, \mathit{model}(X^t(f,k)))$ 
        \EndIf
    \EndFor

    \State $Y^t \gets \textsc{Sim}(D,\{f^t\},\tau,K)$
\EndFor
\end{algorithmic}
\end{algorithm}

\subsubsection{Single-Frame Approach}

The unroll approach solves a single Max-SAT instance over the entire unrolled trace, which can be
expensive for large $K$. 
The single-frame approach instead solves a sequence of much smaller Max-SAT instances, 
one timeframe at a time in a forward manner, 
while re-simulating the full trace after each refinement update.
The core idea of the single-frame approach lies in the observation of how false taints are propagated across timeframes in the design.
False taints persist across time because DFF taints simply forward from one frame to the next.
As a result, a false taint at frame $k$ can be propagated to later frames through tainted state.
The single-frame approach exploits this structure: going forward, at an intermediate frame, 
instead of directly forcing the sink to be
untainted, 
we force the next-state DFF outputs that are falsely tainted to be untainted.
Untainting these falsely tainted DFFs prevents their taints from being propagated into future frames and maintains the invariant that at the current frame, 
there is no incoming false taint from the DFFs,
allowing a single-frame Max-SAT reasoning. 
For a cex,
this approach does guarantee an optimum refinement in the single frame w.r.t. the falsely tainted DFFs but not the optimum refinement to the sink taint,
which might be in several timeframes later.
As will be demonstrated in \cref{sec:evaluation},
with slight optimality loss on the refinement, 
the single-frame approach achieves significant refinement-step speed-up and scalability improvement compared to the unroll approach.

The main algorithm of the single-frame approach is shown in \Cref{alg:single-frame-maxsat}.
At each frame $k$, we form a per-frame candidate set
$\mathcal{C}_k$ and the corresponding fix
variables $\mathcal{R}_k$ (lines 6--7). 
We then construct a per-frame Max-SAT instance $D'_k$
using the same guarded mux encoding as in the unroll approach (via \Call{AddTaintConstraint}{} in
\Cref{alg:add-taint-constraint}), 
but instantiated only for frame $k$ (lines 14--17). 
The key difference from unrolling is
the constraint $\Phi_k$: for intermediate frames ($k<K$), 
we constrain the set of falsely tainted DFF outputs at the
next frame to be cleared, 
i.e., $\Phi_k \equiv (\mathcal{F}_{k+1}=\mathbf{0})$ (lines 8--10). 
For the last frame ($k=K$),
we follow the same constraint for the sink taint $\Phi_K \equiv (y^t_{\text{sink}}(K)=0)$ (line 12) as in the unroll approach.
After solving \Call{MaxSAT}{} on $D'_k$ (line 19), 
we refine the taint logic of the selected blocks using the input-taint
assignment realized in the model at frame $k$ (lines 20--22), 
then re-simulate the full trace under the updated taint logic
(line 23). 
The loop terminates as soon as the sink becomes untainted; otherwise it advances to the next frame. 
Compared to
\Cref{alg:unroll-maxsat}, 
this approach avoids a large $K$-frame Max-SAT instance and is typically more scalable, at the
expense of potentially selecting a non-minimal refinement set.

\section{Implementation}
\label{sec:word-level-cellift}

We instantiate the general CEGAR-T framework on 
\emph{CellIFT-style}~\cite{cellift} instrumentation of RTL designs,
where the taint logic is instrumented for cells (adder, multiplexer, etc.).
The original CellIFT constructs precise taint logic at the \emph{bit-level}, where each signal bit is associated with a taint bit.
In contrast, our prototype adopts a \emph{word-level} variant, 
where each multi-bit RTL word is associated with
one taint bit.
We chose CellIFT as the basis of our prototype because it is a state-of-the-art hardware IFT framework and provides an open-source reference implementation. 
The choice of word-level taint granularity is motivated by Compass~\cite{compass}, which identifies taint-bit granularity as a key design dimension in RTL taint analysis.  
Specifically, RTL designs often declare multi-bit words, which may correspond, for example, to architectural registers or datapath input/output words.
Instead of tracking information flow for each individual bit, we track the information flow of the whole word.
We interpret a word as tainted when at least one of its bits is tainted. 
This granularity reduces the number of taint variables and the size of the instrumented taint logic.
We use the following example to explain the difference between the existing bit-level CellIFT and our word-level CellIFT.

\begin{example}
\label{ex:word-level-taint}

\newcommand{\tvec}[1]{\mathsf{#1}}

Consider a $W$-bit 2:1 mux
\[
\tvec{y} \;=\; s \ternop \tvec{b} \terncol \tvec{a},
\]
where $s\in\{0,1\}$ and $\tvec{a}, \tvec{b}, \tvec{y} \in \{0,1\}^W$.

\paragraph{Precise Bit-level CellIFT}
CellIFT-style bit-level tracking uses one taint bit per data bit.
Let $\tvec{a}^t, \tvec{b}^t, \tvec{y}^t\in\{0,1\}^W, s^t \in \{0, 1\}$ denote the per-bit taint vectors for
$\tvec{a}, \tvec{b}, \tvec{y}, s$, respectively. Then the precise taint logic is
\[
\tvec{y}^t \;=\; \bigl( s \ternop \tvec{b}^t \terncol \tvec{a}^t \bigr)
\;+\; s^{t} \cdot \Bigl( \tvec{a}^t + \tvec{b}^t + (\tvec{a} \oplus \tvec{b} )\Bigr),
\]
where $+$, $\cdot$, and $\oplus$ on vectors are interpreted bit-wise.

\paragraph{Precise Word-level CellIFT}
We use one taint bit per word ($a^t,b^t,y^t$). The precise word-level taint logic is
\[
y^t \;=\; (s \ternop b^t \terncol a^t)
\;+\; s^t \cdot \bigl(a^t + b^t + (\tvec{a} \neq \tvec{b} )\bigr).
\]
\end{example}

For CEGAR-T, word-level granularity is also a practical sweet spot that enables scalable automated refinement and optimization. 
In the general framework, refinement is expressed via an operator $\mathsf{Refine}$ that involves existential quantification. 
Implementing $\mathsf{Refine}$ efficiently for bit-level formulas is nontrivial and often requires additional computation or dedicated data structures ~\cite{bdd-symbolic,jiang2009quantifier} to perform existential quantification.
With a fixed macro-cell library, one alternative is to explicitly construct the refinement lattice for each supported cell type offline and store simplified taint logic at each lattice node, so that online refinement becomes a lightweight hot-swap between precomputed nodes.
The challenge is that the lattice size grows exponentially in the number of input taint bits of the cell. 
Word-level taint bounds this growth by the number of input words of a cell. In our supported library, the maximum number of input words is three (e.g., a mux with two data inputs and one select word).
In fact, the largest refinement lattice for the supported cell types has only eight nodes. 
This makes explicit lattice construction practical and keeps refinement computation lightweight.

We implemented the CEGAR-T prototype in C++.
We use Yosys~\cite{Yosys} for its Verilog front-end and RTLIL IR.
For each refinement step, 
we solve the encoded Max-SAT instances using Z3’s optimization engine~\cite{z3-smt,z3-omt}.


\section{CEGAR-T-based Simulation Worflow}

\label{sec:workflow}

\begin{figure}[t]
    \centering
    \includegraphics[width=\columnwidth]{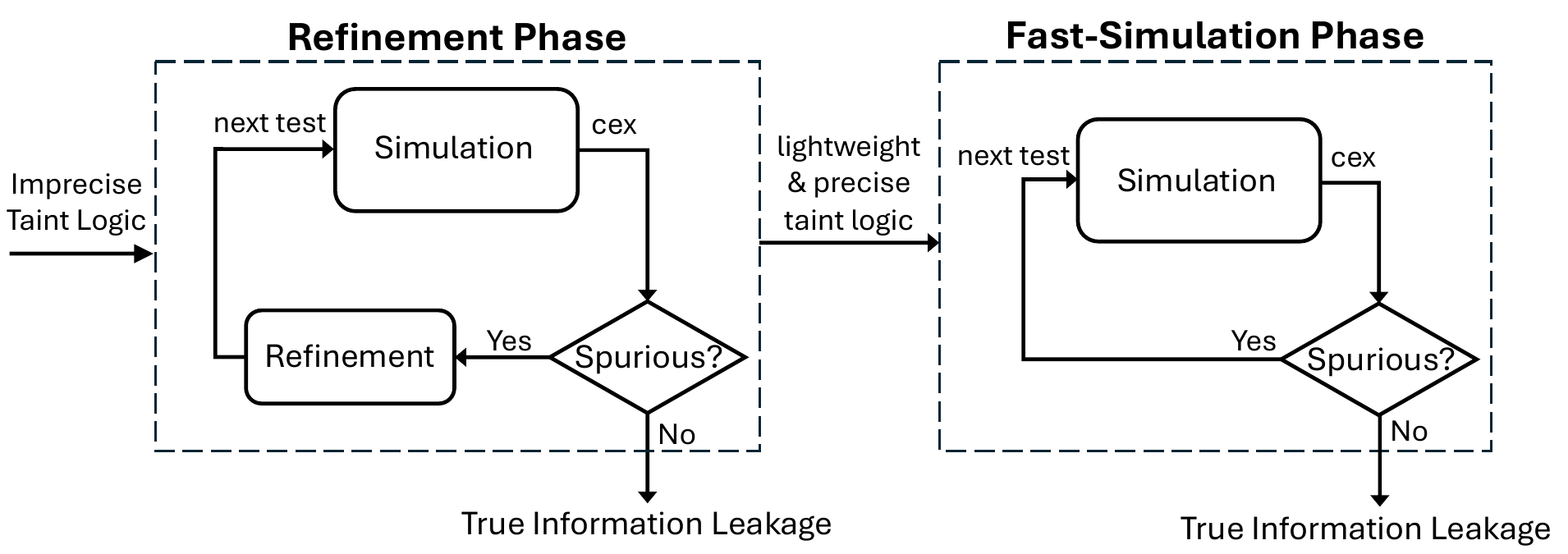}
    \caption{Two-phase simulation-based IFT. It terminates upon detecting true leakage or after exhausting the tests.
    }
    \label{fig:simulation-flow}
\end{figure}

We now explain how CEGAR-T can be used in simulation-based IFT.

Given that we have an automated spuriousness check, each spurious cex does not have to result in a refinement; we could choose to just report it as being spurious and move on to the next simulation test. 
However, if we start with the simplest, most imprecise taint logic, almost all the initial tests will result in spurious cex and the simulation run time will be dominated by the spuriousness check time. 
This motivates our two-phase workflow shown in \Cref{fig:simulation-flow}.
In the Refinement Phase, spurious cex are expected to be frequent so each detected spurious cex is used to refine the taint logic and eliminate a class of similar false positives.
Once the spurious cex are relatively rare and their checking overhead is negligible, 
we transition to the second Fast-Simulation Phase. 
In this Phase, the refined taint logic is fixed and used for large-scale validation. 
Any remaining spurious cex are checked and reported, but they no longer trigger refinement that could slow the simulation throughput.
We use two different simulators in these phases: Icarus Verilog with lower compilation time for the Refinement Phase\footnote{Since a refinement changes the taint logic and requires recompilation, we choose Icarus to minimize the compilation overhead.} and Verilator, which has higher throughput but incurs higher compilation overhead, for the Fast-Simulation Phase.

\section{Evaluation}
\label{sec:evaluation}


We consider the \emph{safe instruction set problem (SISP)}~\cite{conjunct,veloCT}, 
which is proposed for 
specifying the constant-time execution property of a processor on a given \emph{safe instruction set} that consists of constant-time instructions. 
Our benchmarks include two open-source RISC-V in-order cores (1) Ibex~\cite{ibex}, and (2) Rocket~\cite{rocket}, 
as well as the out-of-order one (3) BOOM~\cite{boom-v3} across its Small, Medium, Large, and Mega configurations. 
Following the setup in~\cite{conjunct,veloCT}, 
without modeling memory,
we target \texttt{ibex\_top}, \texttt{Rocket}, and \texttt{BoomCore} modules for Ibex, Rocket, and BOOM variants, respectively.
We model SISP as an information-flow property with the register-file data as the source and the instruction-commit signal (a proxy for timing) as the sink, 
assuming only safe instructions are executed. 

We evaluate CEGAR-T for checking SISP following the workflow in \Cref{sec:workflow}.
Our evaluation asks whether CEGAR-T can derive a lightweight taint logic that preserves SISP-relevant precision while reducing the cost of downstream simulation-based IFT. 
We remark that, for the RTL modules considered, there is no standard benchmark suite, fuzzer, or off-the-shelf testbench tailored to safe-instruction-only validation.
We therefore construct a simulation testbench that generates safe-instruction-only random tests for each evaluated core.
Specifically, the testbench runs $B$ tests, each consisting of $N$ randomly generated safe instructions.

We also applied the workflow to tests that include unsafe instructions. 
In those cases, the spuriousness check detects true information leakage already in the Refinement Phase, causing the workflow to terminate without proceeding to the Fast-Simulation Phase. 
Since our evaluation is intended to measure the simulation-speed benefits of the Fast-Simulation Phase with the optimized taint logic, we focus the reported experiments on safe-instruction only.

All experiments are conducted on a MacOS machine with a 4.5\,GHz Apple M4 Max CPU and 36\,GB memory.

\subsection{Runtime and Taint Logic Reduction Analysis}
\label{sec:phaseI}

\begin{figure}[t]
    \centering
    \includegraphics[width=\columnwidth]{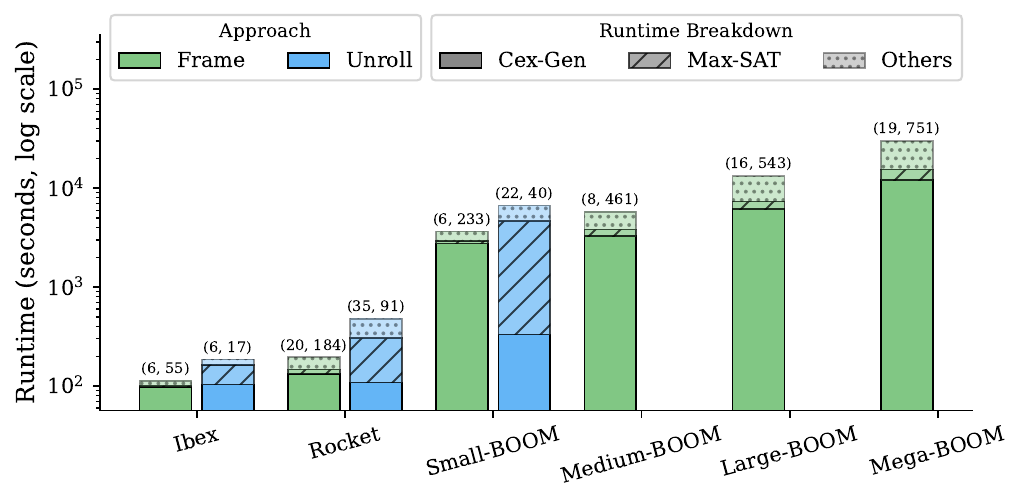}
    \caption{Runtime breakdown of CEGAR-T on different cores. The vertical axis uses log-scale for total runtime, while the breakdown within each bar is drawn linearly to bar length for better visualization. }
    \label{fig:cegar-runtime}
\end{figure}

\begin{figure}[t]
    \centering
    \includegraphics[width=\columnwidth]{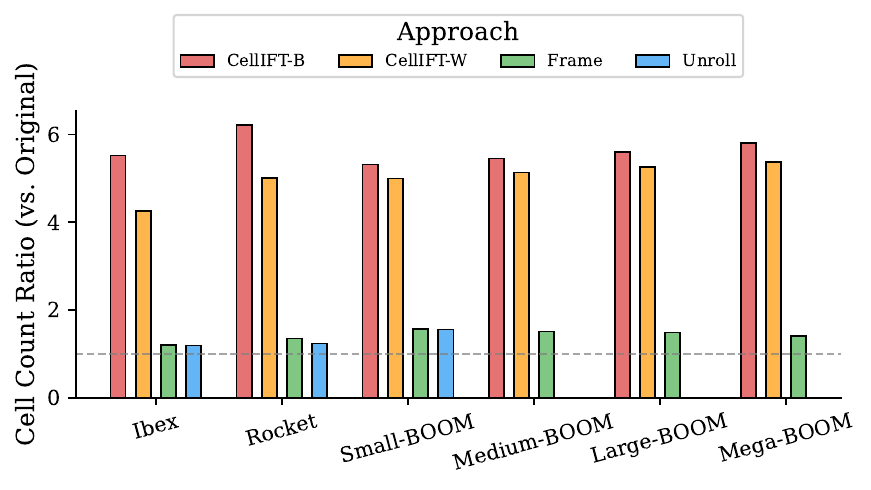}
    \caption{Cell count overhead comparison. The geo-mean cell-count ratios for \texttt{CellIFT-B}, \texttt{CellIFT-W}, \texttt{Frame}, and \texttt{Unroll} are 5.64, 4.99, 1.42, and 1.31, respectively.}
    \label{fig:cell-count-comparison}
\end{figure}

\begin{figure}[t]
    \centering
    \includegraphics[width=\columnwidth]{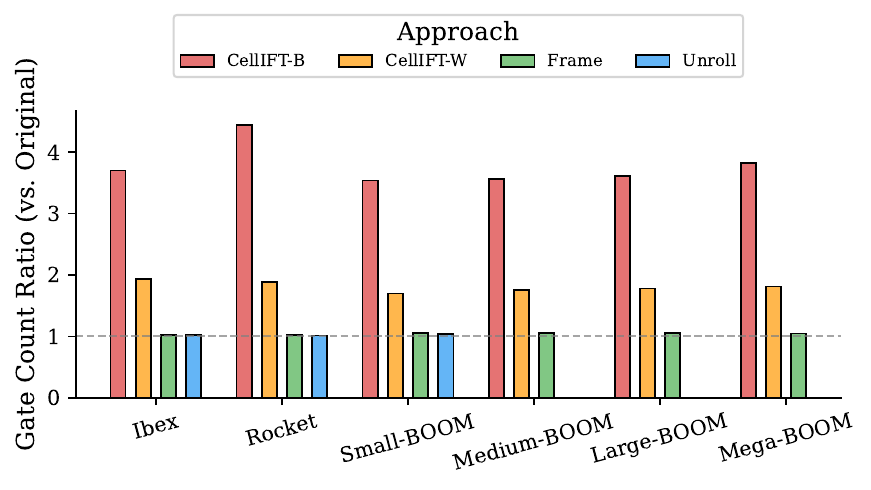}
    \caption{Gate count overhead comparison. The geo-mean of the gate-count ratios for \texttt{CellIFT-B}, \texttt{CellIFT-W}, \texttt{Frame}, and \texttt{Unroll} are 3.77, 1.80, 1.04, and 1.03, respectively.}
    \label{fig:gate-count-comparison}
\end{figure}

We evaluate CEGAR-T using the random safe-instruction testbench with $(N,B)=(20,60K)$.
The parameters are chosen empirically to provide sufficient opportunities to expose spurious cex and drive refinement.
We set a 12-hour timeout for each run.

\Cref{fig:cegar-runtime} shows the runtime breakdown of the refinement process into
cex generation (Cex-Gen), Max-SAT solving (Max-SAT), and other tasks (Others).
The \texttt{Unroll} results of Medium, Large, and Mega BOOM time out and are omitted.
Bar labels (\# iters, \# fixes) denote the number of iterations and refined cells; e.g., (19, 751) for Mega-BOOM \texttt{Frame}.

\texttt{Frame} scales across all evaluated cores, whereas \texttt{Unroll} handles the in-order cores and Small-BOOM but fails to scale to the larger BOOM variants.
This gap is due to the higher Max-SAT cost for the fully unrolled design.
For \texttt{Unroll}, Max-SAT grows substantially with core complexity and becomes a major runtime bottleneck.
For \texttt{Frame}, runtime is dominated by Cex-Gen and Max-SAT remains a small fraction of the total,
showing its effectiveness.
The (\# iters, \# fixes) labels further illustrate the tradeoff between the two refinement algorithms.
\texttt{Unroll} consistently requires fewer refined cells than \texttt{Frame}, consistent with its global optimization over the full unrolled trace.
In contrast, \texttt{Frame} often needs fewer refinement iterations, likely because refining falsely tainted DFFs early in the trace can suppress false taint propagation in later cycles and eliminate additional spurious counterexamples at once.
Together, these results suggest that \texttt{Frame} is the practical scalable choice for refinement, while \texttt{Unroll} can provide stronger per-cex optimization for cases where it does not time out.

We compare the optimized taint logic against both the original bit-level (\texttt{CellIFT-B}) and word-level (\texttt{CellIFT-W}) precise CellIFT baseline.
We implemented \texttt{CellIFT-W} as the corresponding precise word-level taint logic without CEGAR-T optimization
to enable a fair-comparison with CEGAR-T which introduces taint logic at the word level.
To quantify the taint logic reduction achieved, we measure both cell count and gate count after lowering the taint-instrumented design to gate level.
Cell count reflects 
design complexity and correlates with RTL simulation overhead, whereas gate count approximates the cost of physical implementation of the taint logic (e.g., in hardware emulation). 
\Cref{fig:cell-count-comparison} reports the cell-count overhead normalized to the original design without taint logic.
The gate-count results are similar and are shown in \Cref{fig:gate-count-comparison}.
Across both metrics, \texttt{Unroll} and \texttt{Frame} yield substantially smaller taint-logic overhead than both \texttt{CellIFT-B} and \texttt{CellIFT-W}.

\subsection{Downstream Simulation-based IFT}
\label{sec:phaseII}

We next evaluate the end-to-end simulation runtime of the taint logic produced by the Refinement Phase.
In the Fast-Simulation Phase, we use the same workload with $(N,B)=(50,200K)$ to stress the taint logic under substantially larger validation runs.
In addition to \texttt{CellIFT-B} and \texttt{CellIFT-W}, 
we include a baseline (\texttt{No-Refine}) that directly uses the initial imprecise taint logic without refinement.
The \texttt{CellIFT-W} baseline isolates the effect of optimized taint logic on raw simulation runtime, 
whereas \texttt{No-Refine} isolates the cost of low precision in end-to-end validation, 
where tainted traces incur spuriousness-check overhead.

\begin{figure}[t]
    \centering
    \includegraphics[width=\columnwidth]{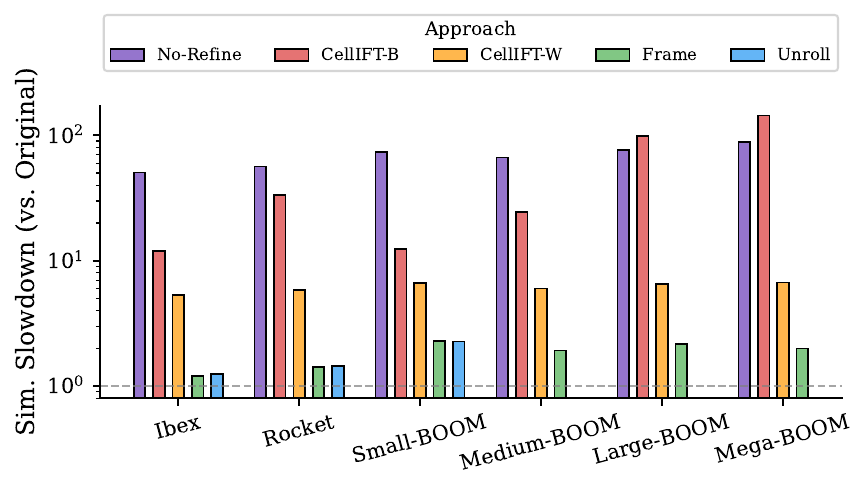}
    \caption{Simulation slowdown with different taint logics. The geo-mean slowdown ratio for \texttt{No-Refine}, \texttt{CellIFT-B}, \texttt{CellIFT-W}, \texttt{Frame}, and \texttt{Unroll} are 67.17, 34.65, 6.16, 1.79, and 1.60, respectively. }
    \label{fig:sim-runtime-comparison}
\end{figure}

\Cref{fig:sim-runtime-comparison} shows the simulation slowdown normalized to the original design.
As the state-of-the-art precise baseline, \texttt{CellIFT-B} incurs huge slowdown across the evaluated cores, especially on the larger BOOM configurations.
The precise word-level baseline \texttt{CellIFT-W} reduces this cost, but still remains substantially slower than the original design.
At the other extreme, \texttt{No-Refine} directly uses the initial imprecise taint logic and is slower than \texttt{CellIFT-W} on every evaluated core, suggesting that the low raw simulation cost of \texttt{No-Refine} can be outweighed by frequent spuriousness checks, 
which account for over 70\% of the total runtime.
This shows that improving precision is also necessary for efficient simulation-based IFT.
In contrast, \texttt{Frame} achieves much lower slowdown by improving both sides of the tradeoff. 
It uses a more lightweight taint logic than the precise CellIFT baselines
and avoids the large spuriousness-check overhead of \texttt{No-Refine}.
\texttt{Unroll} shows results on the cores for which the Refinement Phase completes.

We also measure empirical precision by counting the number of spurious cex.
Both \texttt{CellIFT-B} and \texttt{CellIFT-W} report zero spurious cex on all evaluated cores.
\texttt{Frame} remains very close to these baselines, reporting at most two tainted cex, while \texttt{Unroll} reports at most three on the cores. 
Detailed per-core counts are given in Appendix~\ref{appendix:omitted_fig_tab}. 
Across all the tests, there is no true cex, and therefore all the evaluated cores satisfy the SISP for these tests.  

Although our evaluation focuses on SISP for processors, the same two-phase workflow applies to other information-flow properties and hardware designs. 
Most security properties have a small attacker-observable sink relative to the full design.
Even when the source can influence many internal signals, 
validation only requires sufficient taint precision along the logic relevant to the sink to avoid false positives. 
CEGAR-T exploits this structure by refining only the relavant local taint logic to the propagation of taint reaching sink, 
rather than committing to fully precise ones upfront.
Although the specific refinements may differ across designs and properties, 
we expect similar reductions in instrumentation size and simulation time overhead. 


\subsection{Refinement Analysis: A Case Study}
\label{sec:refinement_case_study}
We analyze the refinements performed by CEGAR-T on Small-BOOM as a case study.
Across the run, refinements primarily tighten how taint propagates through selection logic: the vast majority of refinements target MUX (multiplexer), with only a small number applied to other cells.
This indicates that, in Small-BOOM, 
the main source of imprecision is the overly imprecise taint logic of MUXes, 
which can propagate taint from values on paths not selected by the selecting signal in the MUXes.

\paragraph{Primary refined targets}
In Small-BOOM, the dominant refined blocks fall into three categories:
(i) \emph{register-read and bypass logic}, where operands are selected among regfile read ports, bypass paths, and the constant-zero behavior of \texttt{rf[0]},
(ii) \emph{control/datapath selection cascades}, where downstream control decisions are formed by selecting among multiple candidate conditions and outcomes, and
(iii) \emph{selection logic around data arrays}, where an address-based read is realized as a MUX tree.
In each case, refinements make the taint logic path-sensitive,
i.e., 
a tainted value that is not chosen by a MUX should not taint the output.
This prevents false taints from propagating through tainted state into unrelated control signals and ultimately the sink.

In our run, refinements for categories (i) and (ii) are typically local: each cex triggers a small amount of MUX fixes near the corresponding select logic.
In contrast, refinements for category (iii) can trigger a substantially larger refinement. 
In Small-BOOM, this appears in later iterations with hundreds of MUX fixes.
With an imprecise taint logic in a MUX tree, 
taint from any leaf can taint internal MUX signals even when that leaf is not selected. 
Repeating this across levels can make the tree output tainted whenever any leaf is tainted, regardless of it being selected or not.
Consequently, correcting this behavior requires refining many MUXes in the tree, 
since spurious taint can be introduced locally at many internal MUX outputs.

\paragraph{Limitations of the \texttt{Frame} approach.}
A limitation of the \texttt{Frame} approach is that refinements are driven by removing falsely tainted DFFs in the next frame, 
which can induce extra refinements that are unneeded to remove sink taint.
E.g., Small-BOOM includes debug-related states and memories whose outputs do not functionally affect the instruction commit signal.
Nevertheless, those DFFs can still become falsely tainted due to imprecise taint logic and thus be targeted by refinement.
This can be mitigated by making \texttt{Frame} refinements more property-directed, 
e.g., 
by applying cone-of-influence reduction on the unrolled design so that refinements target DFFs that structurally influence the sink. 

\section{Discussion}
\label{sec:discussion}

\paragraph{Optimality of CEGAR-T Refinements}
The CEGAR-T Unroll and Frame approaches compute optimal refinements only within their respective scopes. 
The Unroll approach minimizes the number of blocks refined for a given cex trace, while the Frame approach minimizes refinements for each time frame within the counterexample. 
As discussed in Section~\ref{sec:phaseI}, Unrolled typically achieves fewer total refinements than Frame, though at greater computational cost for large bounds $K$.
Neither approach guarantees a global, minimal refinement across all counterexamples. 
Determining such a global optimum would require solving a single Max-SAT instance encoding all  observed counterexamples. 
However, this may be computationally prohibitive as it compounds the existing scalability challenges already present in individual unrolled traces for large $K$.

\paragraph{Limitations from word-level taint tracking}
Our instantiation optimizes \emph{word-level} taint logic, 
which is a deliberate design point for managing both taint-logic size and optimization-time. 
However, word-level analysis introduces irreducible imprecision. 
If any bit within a word is influenced by secrets, the entire word is treated as tainted for word-level taint method. 
This can yield spurious propagation in common patterns such as packed fields, masking, bit-slicing, etc. 
As a result, 
while not encountered in our evaluation,
some spurious cex exist only due to word-level coarse granularity. 
Also, the learned refinements may compensate for word-level over-approximation rather than capturing the bit-precise flow. 
Despite this, our word-level approach remains valuable as it systematically removes spurious taint for a given security property while preserving scalability. 

\section{Related Work}
\label{sec:related-work}

\paragraph{Hardware Information Flow Tracking} 
Prior work on hardware IFT has explored self-composition, taint analysis, as well as hybrid approaches that combine both methods. 
The self-composition~\cite{self-composition} and, taint analysis approaches at different abstraction levels~\cite{glift,rtlift,cellift},
and their optimization~\cite{impreciseIFT2016,impreciseIFT2017,hu2018property,compass} are discussed in \cref{sec:introduction}.

\paragraph{Secure Processor Verification Frameworks}

Prior work on secure processor verification has largely been developed in self-composition-based formal verification settings.
One line of work studies secure speculation through hardware--software contracts~\cite{guarnieri2021hardware} using frameworks such as~\cite{upec-2023,leave,contract-shadow-logic}.
These contract-based properties are outside the scope of our evaluation because they cannot be expressed directly as a single source-to-sink information-flow property.
Another line of work studies constant-time processor verification through the safe instruction set problem (SISP)~\cite{conjunct,veloCT} with self-composition and invariant synthesis.
In contrast, our work also targets processor verification under SISP, but focuses on simulation-based IFT rather than self-composition-based formal verification.


\paragraph{Max-SAT for Design Debugging and Diagnosis.}

Max-SAT has been used extensively in VLSI debugging/diagnosis to explain failing traces by selecting a minimum set of suspect fault sites~\cite{safarpour_fmcad07,chen_glsvlsi09,chen_tcad10}.
We use Max-SAT inside a CEGAR loop to select a minimum-cost taint-logic refinements that eliminate spurious information-flow counterexamples, 
rather than to localize functional bugs.

\section{Conclusion}
\label{sec:conclusion}

We presented \emph{CEGAR-T}, a fully automated framework for property-driven taint-logic optimization.
The core contribution of CEGAR-T is a principled and automated refinement step that turns taint-logic optimization from a heuristic, human-driven process into a systematic optimization framework.
To this end, we formalized taint-logic precision using non-interference semantics, characterized the refinement space through a lattice-like structure, and developed an automated refinement procedure that optimizes both \emph{how} to refine, via the refinement lattice, and \emph{where} to refine, via Max-SAT.
We further introduced two complementary refinement algorithms, the Unroll and the Single-Frame approaches, that offer different optimality--scalability tradeoffs.
Our evaluation demonstrates that CEGAR-T reduces taint logic overhead by 91.32\%/89.88\% compared to bit-level CellIFT and its word-level variant, 
and achieves up to 19.35$\times$/3.44$\times$ corresponding speedups in simulation-based IFT.
By significantly reducing the simulation overhead, CEGAR-T advances the scalability of 
taint-based security verification, opening the door for full industrial adoption.
\bibliographystyle{IEEEtran}
\bibliography{ref}
%



\clearpage

\appendices

\section{Additional Figures and Tables for Evaluation}
\label{appendix:omitted_fig_tab}

\Cref{tab:phase2-precision} shows the number of spurious cex out of 200K.
\texttt{Frame} reports at most two tainted cex, while \texttt{Unroll} reports at most three on the evaluated cores.
This suggests high empirical precision of the synthesized lightweight taint logic.

\begin{table}[h]
\centering
\caption{Number of spurious cex out of 200K validation tests.}
\label{tab:phase2-precision}
\begin{tabular}{lcc}
\toprule
Core & \texttt{Frame} & \texttt{Unroll} \\
\midrule
Ibex         & 0 & 0 \\
Rocket       & 2 & 2 \\
Small-BOOM   & 0 & 3 \\
Medium-BOOM  & 1 & -- \\
Large-BOOM   & 2 & -- \\
Mega-BOOM    & 2 & -- \\
\bottomrule
\end{tabular}
\end{table}

\begin{figure}[h]
    \includegraphics[width=\columnwidth]{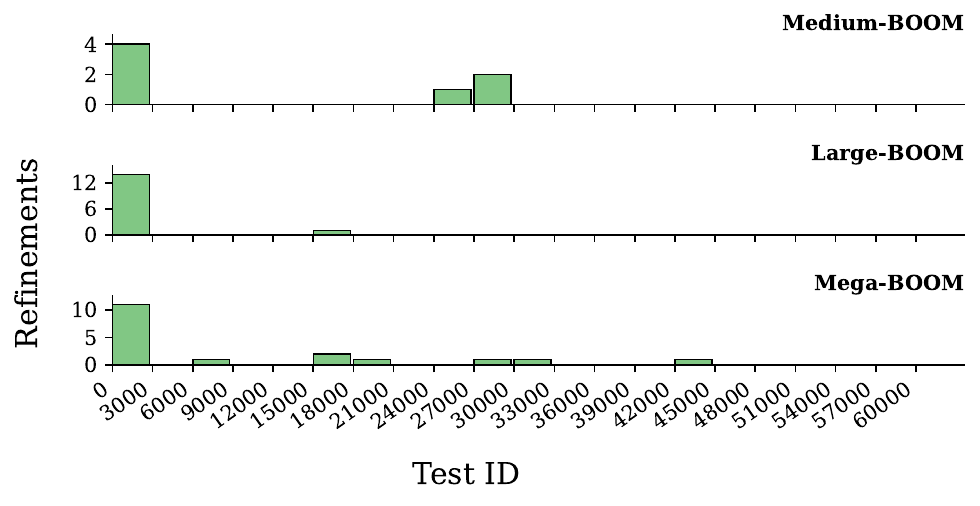}
    \caption{
    Number of refinements per 3000-test bucket during Phase~I for three representative BOOM variants.
    The ID is indexed in the sequence of when the test is sequentially simulated.
    }
    \label{fig:refinement_rate}
\end{figure}

\Cref{fig:refinement_rate} reports the number of refinements in each 3000-test bucket for three representative large BOOM configurations: Medium, Large, and Mega.
Across all three designs, refinements are strongly front-loaded, with the first bucket containing the majority of refinements.
After this initial burst, the number of refinements drops sharply, but a sparse long tail remains, with occasional later buckets still triggering additional updates.
This pattern indicates that the Refinement Phase quickly removes many easy-to-expose sources of imprecision, while some spurious counterexamples continue to arise later in the test sequence.
We observe similar behavior on the other evaluated cores.
Overall, this front-loaded refinement pattern matches the intended two-phase workflow. 
Early traces primarily drive refinement, while later traces trigger far fewer updates and motivate freezing the resulting taint logic for the Fast-Simulation.

\end{document}